\magnification=1200
\baselineskip 14pt
\font \bigbf = cmbx10 scaled\magstep2

\def\vf{\varphi}

\def\f{f_\kappa}
\def\com{\; , \;}

\def\bo{\otimes}

\def\tks{{\widetilde K}_{P_1} \wedge \cdots \wedge {\widetilde K}_{P_s}}
\def\as{A_{P_1} \wedge \cdots \wedge A_{P_s}}

\def\asi{A_{P_1} \wedge \cdots \wedge D_{k+1}A_{P_i} \wedge \cdots \wedge 
A_{P_s}}
\def\sp{\medskip}
\def\L{\Lambda}

\def\cwt{\widetilde{\cal W}}
\def\cyt{\widetilde{\cal Y}}

\def\ct{\widetilde{C}}
\def\ni{\noindent}

\def\bp{{\bf  P}}
\def\br{{\bf R}}
\def\bn{{\bf N}}
\def\bc{{\bf C}}
\def\ss{\subset}
\def\ct{{\cal T}}
\def\cp{{\cal P}}
\def\ch{{\cal H}}
\def\cc{{\cal C}}

\def\n{\Vert}

\vskip .5in
\centerline{{\bigbf AN ALTERNATIVE TO }}
\vskip .1in
\centerline{{\bigbf PLEMELJ-SMITHIES FORMULAS ON INFINITE }}
\vskip .1in
\centerline{{\bigbf DETERMINANTS}}

\vskip .5in
\centerline{{\bf Domingos H. U. Marchetti }}
\centerline{Department of Mathematics and Statistics}
\centerline{McMaster University }
\centerline{Hamilton, Ontario L8S 4K1, Canada }
\sp
\ni

\vskip 1in
\ni
{\bf ABSTRACT}: An alternative to Plemelj - Smithies formulas for
the p -regularized quantities $d^{(p)}(K)$ and $D^{(p)}(K)$ is presented 
which generalizes previous expressions with $p=1$ due to Grothendieck and 
Fredholm. It is also presented global upper bounds for these quantities. In 
particular we prove that
$$
|d^{(p)}(K)| \le e^{\kappa \n K \n_p^p}
$$
holds with $\kappa = \kappa(p) \le \kappa(\infty) = \exp \bigl\{-{1 \over 
4(1+ e^{2 \pi})} \bigr\}$
for $ p\ge 3$ which improves previous estimate yielding $\kappa(p) = e
(2 + \ln(p-1))$.

\vskip .2in

\ni
{\it to appear in J. Funct. Analysis}

\vskip .1in
\ni


\par\vfil\eject

\ni
{\bf 1.  INTRODUCTION}
\sp

The Fredholm Theory$^{[1]}$, as addressed in this paper,
is concerned with the problem of solving the equation
$$
(1 + \mu \; K) f = f_0    \eqno (1.1)
$$
in a separable Hilbert space $\ch$ with $K$ belonging to the trace - class 
of operators on $\ch$ or, more generally$^{[2,3,4,5]}$, $K$ is considered 
to be a compact operator of the class ${\cc}_p = \{A \; : \; \n A \n^p_p 
\equiv Tr( |A|^p)
< \infty \}$. Fredholm theory has been mainly applied in 
Scattering Theory
but (1.1) also appears in a variety of problems in Many Body Theory
and Quantum Field Theory 
(see Simon$^{[6,7]}$ for a review and applications).

The basic result of the Fredholm Theory is to write the resolvent 
$R(K;\mu) \equiv (1 + \mu K)^{-1}$ of the operator $K$ as a quotient
$$
R(K;\mu) = {D^{(p)}(K;\mu) \over d^{(p)}(K;\mu)}, \;\;\;\;\;\;\;\;\;\;\;
p = 1, 2, \dots  \eqno (1.2)
$$
of entire functions of $\mu$. Therefore, (1.1) has a unique solution
for any operator of the class ${\cc}_p$, given by
$$
f = R(K;\mu)\; f_0   \eqno (1.3)
$$
provided $-1/\mu$ is not an eigenvalue of $K$.

$d^{(p)}$ and $D^{(p)}$ can be explicitly written in terms of their power
series
$$
\eqalignno{
d^{(p)}(K;\mu) &= \sum_{n \ge 0} \; {\mu^n \over n!} \; d_n^{(p)}(K) & (1.4)
\cr
D^{(p)}(K;\mu) &= \sum_{n \ge 0} \; {\mu^n \over n!} \; D_n^{(p)}(K) & (1.5)
\cr}
$$
where $d^{(p)}_n$ are complex valued and $D^{(p)}_n$ are operator valued 
coefficients given by the Plemelj$^{[3]}$ - Smithies$^{[4]}$ formulas
$$
d_n^{(p)}(K) = \left|\matrix{
\sigma_1 & n-1 & 0 & \ldots & 0 \cr
\sigma_2 & \sigma_1 & n-2 & \ldots & 0 \cr
\sigma_3 & \sigma_2 & \sigma_1 & \ldots & 0 \cr
\vdots & \vdots & \vdots & \ddots & \vdots \cr
\sigma_n & \sigma_{n-1} & \sigma_{n-2} & \ldots & \sigma_1 \cr }\right|
\eqno (1.6)
$$
and
$$
D^{(p)}_n(K) = \left|\matrix{
K^0 & n & 0 & \ldots & 0 \cr
K^1 & \sigma_1 & n-1 & \ldots & 0 \cr
K^2 & \sigma_2 & \sigma_1 & \ldots & 0 \cr
\vdots & \vdots & \vdots & \ddots & \vdots \cr
K^n & \sigma_n & \sigma_{n-1} & \ldots & \sigma_1 \cr}\right|  \eqno(1.7)
$$
with $K^0 \equiv I$ and
$$
\sigma_j = \sigma_j(p) = \cases{{\rm Tr}(K^j) \;\;\;\;\;\;\;\; & if 
$\;\;\; j \ge p$ 
\cr
0  \;\;\;\;\;\;\;\; & otherwise \cr} \eqno (1.8)
$$




Expression (1.5) is to be interpreted in the sense that $(\phi, D^{(p)}_n(K)
\psi)$ is the determinant of the matrix 
(1.7) with the operator $K^i$ replaced
by $(\varphi, K^i \; \psi)$, for $\varphi, \psi \in \ch$.


The function $d^{(p)}$ is called the p - regularized determinant and we 
write here Poincar\'{e}'s definition$^{[2]}$
$$
d^{(p)}(K;\mu) = {\rm det}_p (1 + \mu \; K) \equiv \exp 
\biggl\{\sum_{j=p}^\infty (-1)^{j+1}
{\mu^j \over j} {\rm Tr} \; (K^j) \biggr\} \eqno (1.9)
$$

Although definition (1.9) seems to
require $ \mu \n K \n_p <1$ (see Simon$^{[6]}$ for other definition),
it can be used to derive (1.6) for $\mu$ sufficiently small
and by Hadamard's inequality$^{[4]}$, or by a limite procedure$^{[5]}$,
one can prove analyticity of (1.4)
for all $\mu \in {\bf C}$.

In this note (1.9) is used to
derive an alternative formula for $d^{(p)}$ and $D^{(p)}$,  with $p=1,
2, \dots$, which generalize  the algebraic formula for the determinant
$d^{(1)}$ due to Grothendieck$^{[9]}$. We then discuss the analytic properties
of these series.

Grothendieck's formula is given by the power series (1.4) where the 
n-th coefficient
$$
d^{(1)}_n = n! \; {\rm Tr}(\wedge^n K) \equiv {\ct}_n(\wedge^n K) \eqno (1.10)
$$
is a trace of a n-fold antisymmetric tensor product of operators on $\ch$ and,
as recognized by Simon$^{[6]}$, it
has the following advantage: once one
uses the simple bound
$$
| {\ct}_n(\wedge^n A)| \le \n A \n^n_1 \eqno (1.11)
$$
the analytic properties of $d^{(1)}$ are established avoiding to use  
Hadamard's inequality. It is also worth of mention that (1.10) reduces to the
definition of Fredholm$^{[1]}$ if $K$ is an integral operator with continuous
kernel.

We shall state our results.

\proclaim Theorem 1.1. Let $K \in {\cc}_p$. Then the 
power series (1.4) and (1.5) converge for all $\mu \in \bc$ and their
coefficients can be written as
$$
\eqalignno{
d^{(p)}_n & = \sum_{\bp \in {\cp}_n} \left[ \prod_{i=1}^s C^{(p)}_{|P_i|}
\right] {\ct}_s(K^{|P_1|} \wedge \dots \wedge K^{|P_s|}) & (1.12) 
\cr
D^{(p)}_n & = {1\over n+1} \sum_{\bp \in {\cp}_{n+1}} 
(-1)^{|P_1|-1} |P_1| ! \left[ \prod_{i=2}^s C^{(p)}_{|P_i|}\right] 
K^{|P_1|-1} \; {\ct}_{s-1}(K^{|P_2|} \wedge \dots \wedge K^{|P_s|}) & \cr
&  & (1.13) \cr} 
$$
where ${\cp}_n$ is the collection of partitions $\bp = (P_1, \dots, P_s)$
of $\{1, 2, \dots, n\}$ and
$$
C^{(p)}_k = \left[ \left( {d \over d\lambda} + \xi_p \right)^k (1+ \lambda) 
\right]_{\lambda = 0} \eqno (1.14)
$$
with $\xi_p = \xi_p(\lambda)$ such that $\xi_1 = 0$ and
for $p  \ge 2$
$$
\xi_p = \sum_{j=0}^{p-2} (-1)^{j+1} \; \lambda^j  \eqno (1.15)
$$

\sp
\proclaim Remark 1.2. One can check from (1.14) that
$$
C^{(p)}_k = 0 \;\;\;\;\;\;\;\; {\rm for} \;\;\;\; k = 1, \dots, p-1. 
\eqno (1.16)
$$
Therefore (1.12) and (1.13) are well defined expressions since 
$$
|\ct_s(K^{n_1} \wedge \cdots \wedge K^{n_s})| \le 
\prod_i \n K \n^{n_i}_{n_i}   \eqno (1.17)
$$
is finite provided $n_1, \dots, n_s \ge p$. 
So, the effect of (1.16) in (1.12) is analogous to that of (1.8) in
the Plemelj - Smithies formula of $d^{(p)}_n$ (recall that $|\sigma_q
(p)|^{1/q} \le \n K \n_q \le \n K \n_p$ for any $q \ge p$).

\sp
\proclaim Remark 1.3.  Analyticity of (1.4) and (1.5) can be
established by estimating (1.12) and (1.13). If the following crude bound
for (1.14) 
$$
|C^{(p)}_k| \le 2 \; p^{k-1} \left( {p-2 \over p-1} k \right) !  \eqno (1.18)
$$
is used one can show covergence of these series for all $\mu \in \bc$.
In (1.18) $2p^{k-1}$ accounts for the number of terms after expanding (1.14)
and we then take the worse of these. It is not difficult to provide an
upper bound on $C_k^{(p)}$ which replaces $2 p^{k-1}$ in (1.18) by a $p$
independent constant $c$.

\sp
A different expression for (1.9) is needed in order to obtain further analytic
properties. By Lidskii's theorem$^{[13,6]}$ (1.9) can be written as
$$
d^{(p)}(K;\mu) = \prod_{i \in I} E(-\mu \gamma_i ; p-1) \eqno (1.19)
$$
where $E(z;q)$ is the Weierstrass primary factor defined by
$$
E(z;0) = 1-z  \eqno (1.20)
$$
and 
$$
E(z;q) = (1-z) \; \exp \biggl\{\sum_{j=1}^q {z^j \over j} \biggr\} 
\eqno (1.21)
$$
for $q > 0$. Here $\{\gamma_i\}_{i \in I}$ is the collection of all 
eigenvalues of $K$, counted up to algebraic multiplicity. 

By an improved estimate on the Weierstrass factor $E$ (Lemma 3.1.) which 
sharpens previous bound on its type$^{[12]}$, we are led to the following 
result:

\sp
\proclaim Theorem 1.4. Given $p \ge 1$, let $K \in \cc_p$. The following 
inequalities hold
$$
|d^{(p)}(K;1)| \le e^{\kappa \n K \n^p_p}  \eqno (1.22)
$$
and
$$
\n D^{(p)}(K;1) \n_\infty \le 2 \; e^{\kappa (1 + \n K \n_p)^p}  \eqno (1.23)
$$
with $\kappa =  \kappa(p)$ such that $\kappa(1) =1 \com \kappa(2) = {1
\over 2}$ and for $p \ge 3$
$$
\kappa =  {p-1 \over p} \exp \biggl\{- {p-2 \over 4 p} \biggl[ 1+ \bigg( 1+ 
2 \biggl(1+ {\rm cosec} {\pi \over p} \biggr)^{-1} \biggr)^{p-1} 
\biggr]^{-1} \biggr\} \; . \eqno (1.24)
$$

\sp
\proclaim Remark 1.5. The proof of Theorem 1.4 is straightforward for $p=1$; 
Smithies$^{[4]}$ established the result for $p=2$ and for $p=4$ 
Brascamp$^{[5]}$ obtained $\kappa = {3 \over 4}$. Our proof of theorem 1.4
extends for arbitrary $p$ the proof of ref. [5]. Notice that $\kappa(p) \le
\kappa(\infty) = e^{{-1\over 4(1 + e^{2\pi})}}$ is in contrast
with the previous estimate$^{[12,6]}$ resulting in  
$\kappa(p) = e(2+ \ln (p-1))$.

\sp

Theorem 1.1 and theorem 1.4 will be proved in section 3. Theorem 1.1 is based 
on an explicit algebraic computation of the n-th  derivative of $d^{(p)}$ 
(Lemma 2.3 of section 2). The expression derived in lemma 2.3 is important 
for organizing terms in the cluster expansion of fermionic Quantum
Field Theories$^{[10]}$. Applications on this field will appear elsewhere$^{[
11]}$.
In section 4 it is presented a simple derivation of Fredholm's formula
and its generalization.

\vskip 1in
\eject 

\ni
{\bf 2. THE BASIC LEMMA}
\sp

We begin with a brief review on the antisymmetric tensor product.
We use the notation and some results of ref. [10].

Let $\otimes^n \ch = \ch \otimes \cdots \otimes \ch$
be the n - fold tensor product of a Hilbert space
$\ch$ and let $\wedge^n \ch = \ch \wedge \cdots \wedge \ch$ donote its
antisymmetric subspace. A "simple" vector $\Phi \in \wedge^n \ch$ is of the
form
$$
\eqalignno{
\Phi & = {1 \over n !} \sum_\pi (-1)^{|\pi|} \varphi_{\pi(1)}
\bo \vf_{\pi(2)} \bo \cdots \bo \vf_{\pi(n)}  &
\cr
& \equiv \Pi (\vf_1 \bo \cdots \bo \vf_n)  & (2.1) \cr}
$$
for some $\vf_1, \dots, \vf_n \in \ch$. Here we sum over all permutations
$\pi = (\pi(1), \dots, \pi(n))$ of $\{1, 2, \dots, n\}$ and $|\pi|$
counts the number of permutations required to return to the original 
order. $\Pi$ stands for the projection of $\bo^n \ch$ into $\wedge^n \ch$.
We write $\Phi = \vf_1 \wedge \cdots \wedge \vf_n$.

If $\{\vf_i \}$ is an orthonormal basis for $\ch$ then $\{\vf_{i_1} \wedge
\vf_{i_2} \wedge \cdots \wedge \vf_{i_r} \}$ with $i_1 < i_2 < \cdots 
< i_r$  is an orthonormal basis for $\wedge^r \ch$,
$r=1, 2, \dots$ . From this (1.11) and its generalization (1.17) 
can be proved.

If $\Phi\com \Psi \in \wedge^n \ch$ are "simple" vectors, 
their scalar product is given by the
determinant of a $n\times n$ matrix
$$
(\Phi \com \Psi) = {1 \over n!} \det \{ (\phi_i, \psi_j) \} \eqno (2.2)
$$
whose elements are scalar products in $\ch$.

Given $K_1, \dots, K_n$ bounded operators on $\ch$ and $\Phi \in \wedge^n \ch$
we define 
$$
(K_1 \wedge \cdots \wedge K_n) \Phi = {1 \over n !} \sum_\pi K_{\pi(1)}
\vf_1 \wedge \cdots \wedge K_{\pi(n)} \vf_n  \eqno (2.3)
$$
i.e. $K_1 \wedge \cdots \wedge K_n = \Pi (K_1 \bo \cdots \bo K_n) \Pi$.
We write $\wedge^n K = K\wedge \cdots \wedge K$.

If $K$ and $L$ are operators on $\wedge^n \ch$ and $\wedge^m \ch$, 
respectively,
then $K \wedge L = \Pi (K \bo L) \Pi$ is an operator in $\wedge^{n+m} \ch$.
Moreover the product $\wedge$ is commutative, associative and distributive
with respect to addition. 


Given a bounded operator $K$ on $\ch$, we define its derivation
$d(\wedge^n K)$ on $\wedge^n \ch$ by
$$
d(\wedge^n K) = n (K \wedge I \wedge \cdots \wedge I)  \eqno (2.4)
$$

\sp
\proclaim Lemma 2.1. Let $K_1, \dots, K_n$ and $L$ be bounded operators 
on $\ch$. Then
$$
\eqalignno{
\wedge^n L \cdot K_1 \wedge \cdots \wedge K_n  & = L K_1 \wedge \cdots
\wedge L K_n  & (2.5)
\cr
d\wedge^n L \cdot K_1 \wedge \cdots \wedge K_n  & =  \sum_{i=1}^n 
K_1 \wedge \cdots \wedge L K_i \wedge \cdots \wedge K_n  & (2.6)
\cr}
$$

\sp
\ni
{\it{\bf proof:}} Appendix of [10].

\sp
\proclaim Lemma 2.2$^{[10]}$. Let $A_1, \dots, A_{k+1}$ be trace class
operators on $\ch$. Then
$$
\ct_{k+1}(A_1 \wedge \cdots \wedge A_{k+1}) = \ct_1(A_{k+1}) \ct_k
(A_1 \wedge \cdots \wedge A_k) - \ct_k(d\wedge^k A_{k+1} \cdot A_1 
\wedge \cdots \wedge A_k) \eqno (2.7)
$$

\sp
\ni
{\it{\bf proof:}} We notice that (1.10) implies 
$$
\left[ \left(\prod_{i=1}^{k+1} {d \over d\lambda_i} \right)  \det(1+A(\lambda))
\right]_{\lambda=0} = \ct_{k+1}(A_1 \wedge \cdots \wedge A_{k+1}) \eqno (2.8)
$$
where $A(\lambda) = \lambda_1 A_1 + \cdots + \lambda_{k+1} A_{k+1}$. We write
$$
\det(1 + A(\lambda)) = \det(1 + \lambda_{k+1} A_{k+1}) \det(1 + R_{k+1}
A({\tilde \lambda})) \eqno (2.9)
$$
where $R_{k+1} = R(A_{k+1};\lambda_{k+1})$ and ${\tilde \lambda} = (\lambda_1,
\dots, \lambda_k)$.

>From (2.5), (2.8) and (2.9) we have
$$
\left[ \left(\prod_{i=1}^{k} {d \over d\lambda_i} \right)  \det(1+A(\lambda))
\right]_{{\tilde \lambda}=0} = \det(1 + \lambda_{k+1} A_{k+1})
\ct_{k}(\wedge^k R_{k+1} \cdot A_1 \wedge \cdots \wedge A_{k}) \eqno (2.10)
$$

We deduce (2.7) by differentiating (2.10) with respect to $\lambda_{k+1}$
and setting $\lambda_{k+1} = 0$.

\sp
We are now ready to state our basic lemma.

\sp
\proclaim Lemma 2.3. Let $A \in \cc_p$ be a multivariable function 
and let $D_1, \dots, D_n$ be derivatives. Then we have
$$
\biggl( \prod_{j=1}^n D_j \biggr) d^{(p)}(A;1) = \sum _{\bp \in \cp_n}
\ct_s(\wedge^s R \cdot A_{P_1} \wedge \cdots \wedge A_{P_s}) \; d^{(p)}
(A;1) \eqno (2.11)
$$
where $R= R(A;1)= (1 + A)^{-1}$ and for any subset $Q$ of $\{1, \dots, n\}$
$$
A_Q = \biggl(\prod_{j \in Q} (D_j + \xi_p D_j A) \biggr) (1 + A) \eqno (2.12)
$$
with $\xi_p = \xi_p(A)$ as in (1.15) with $\lambda$ replaced by $A$.

\sp
\proclaim Remark 2.4. The cancelation leading to (1.16) occurs for (2.12):
terms in the expansion of $R \; A_Q$ are of the form
$$
A^q D_{Q_1}A \dots D_{Q_n}A
$$
for a $q \in \bn$ and a partition $(Q_1, \dots, Q_n)$ of $Q$ such that 
$q+n \ge p$. Here $D_{Q_i} = \prod_{k \in Q_i} D_k$. So, each monomial in 
$A$ and/or derivatives of $A$ has at least order $p$. We are assuming that all
these terms are in $\cc_1$.

\sp
\ni
{\it {\bf proof:}} We prove Lemma 2.3 by induction. We write $d^{(p)}= d^{(p)}
(A;1)$.

\sp
\ni
{\it First step:} Differentiating once (1.9) gives
$$
D_1 d^{(p)} = \sum_{j=p}^\infty (-1)^{j-1} \; 
\ct_1(A^{j-1} D_1 A) \; d^{(p)} \eqno (2.13)
$$

Since
$$
\eqalignno{
\sum_{j=p}^\infty (-1)^{j-1} A^{j-1} D_1A &= (R + \xi_p)D_1 A & 
\cr
&= R [D_1 + \xi_p D_1 A] (1 + A) & (2.14) \cr}
$$
we have from (2.12)
$$
D_1 d^{(p)} = \ct_1(R \;A_{\{1\}}) \; d^{(p)}  \eqno (2.15)
$$
which establishes (2.11) for $n=1$.

\sp 
\ni
{\it Induction step:} We now assume (2.11) valid for $n=k$. By differentiating
(2.11) with respect to $(k+1)$-th variable and using (2.15) with $\{1\}$
replaced by $\{k+1\}$ it follows that
$$
\eqalignno{
& \biggl( \prod_{j=1}^{k+1}  D_j  \biggr) d^{(p)} &
\cr
& =  \sum_{\bp \in \cp_k}
\left\{ D_{k+1} \ct_s(\wedge^s R \cdot \as) + 
\ct_s(\wedge^s R \cdot \as)
\ct_1(R A_{\{k+1\}}) \right\} d^{(p)} & 
\cr 
&   & (2.16) \cr}
$$

We have
$$
\eqalignno{
D_{k+1} \ct_s(\wedge^s R \cdot  \as) 
& =  \sum_{i=1}^s \ct_s(\wedge^s R \cdot \asi) &
\cr
& - \ct_s(\wedge^s R \cdot d\wedge^s(R D_{k+1} A) \cdot \as)  &
(2.17)  \cr}
$$

The second term in the right hand side of (2.17) can be written as
$$
\ct_s (\wedge^s R \cdot d\wedge^s(\xi_p D_{k+1} A) \cdot \as) 
- \ct_s(\wedge^s R \cdot d\wedge(R A_{\{k+1\}}) \cdot \as) \eqno (2.18)
$$

Now, it follows from (2.6) that
$$
\eqalignno{
\sum _i \asi & + d\wedge^s(\xi_p D_{k+1} A) \cdot \as   & 
\cr
= \sum_i A_{P_1} \wedge \cdots \wedge & (D_{k+1} + \xi_p D_{k+1} A) A_{P_i}
\wedge \cdots \wedge A_{P_s}) & 
\cr
= \sum_i A_{P_1} \wedge \cdots \wedge & A_{P_i \cup \{k+1\}} \wedge \cdots
\wedge A_{P_s} & (2.19) \cr}
$$

>From (2.16) - (2.19) we have
$$
\biggl( \prod_{j=1}^{k+1} D_j \biggr) d^{(p)} = \sum_{\bp \in \cp_k}
\left({\cal W}_{\bp} + {\cal Y}_{\bp} \right) d^{(p)} \eqno (2.20)
$$
where
$$
{\cal W}_{\bp}  = \ct_s(\wedge^s R \cdot \as) \ct_1(R A_{\{k+1\}}) 
- \ct_s(\wedge^s R \cdot d\wedge^s(R A_{\{k+1\}}) \cdot \as)
$$
and
$$
{\cal Y}_{\bp} = \sum_{i=1}^s \ct_s(\wedge^s R \cdot A_{P_1} \wedge
\cdots \wedge A_{P_i \cup \{k+1\}} \wedge \cdots \wedge A_{P_s})
\eqno (2.21)
$$

We now set in Lemma 2.2 $A_j = R A_{P_j}$ for $j=1, \dots, k$ and $A_{k+1}
= R A_{\{k+1\}}$. It follows from Lemma 2.1, Lemma 2.2 and Remark 2.4 that
$$
{\cal W}_{\bp}  =
\ct_{s+1}(\wedge^{s+1} R \cdot \as \wedge
A_{\{k+1\}})  \eqno (2.22)
$$



>From (2.21) and (2.22) we can write (2.20) as
$$ 
\sum_{\bp \in \cp_{k+1}} \ct_s(\wedge^s R \cdot \as) d^{(p)} \eqno (2.23) 
$$
which proves that (2.11) is also valid for $n=k+1$, completes our induction
argument and proves Lemma 2.3.

\vskip 1in
\eject

\ni
{\bf 3. PROOF OF THEOREMS 1.1 AND 1.4}

\sp

Let $D_1 = \dots = D_n = {d \over d\lambda}$ and $A = \lambda K$ in Lemma 2.4.
Since we have for any $Q \subset \{1, \dots, n\}$
$$
\eqalignno{
\left[ K_Q \right]_{\lambda=0} & = \left[ \left( {d \over d\lambda} 
+ \xi_p(\lambda K) K 
\right)^{|Q|} (1 + \lambda K) \right]_{\lambda =0} &
\cr
& = \left[ \left( {d \over d\lambda} 
+ \xi_p(\lambda)  
\right)^{|Q|} (1 + \lambda) \right]_{\lambda =0} K^{|Q|} & (3.1) \cr}
$$
(1.12) follows from (2.11).

Let us assume that $R(K;\lambda) d^{(p)}(K;\lambda)$ is an analytic (entire)
operator valued function of $\lambda$. We have from (1.2) that
$$
\eqalignno{
D^{(p)}_n(K) & = \left[ \left( {d \over d\lambda} \right)^n \left( R(K;\lambda)
d^{(p)}(K;\lambda) \right) \right]_{\lambda=0} &
\cr
& = {1\over n+1}\sum_{m=1}^{n+1} {n+1 \choose m} (-1)^{m-1} m! \; 
K^{m-1} \sum_{\bp \in \cp_{n+1-m}} 
\left( \prod_i C^{(p)}_{|P_i|} \right) \ct_s(K^{|P_1|} \wedge \cdots \wedge
K^{|P_s|}) & \cr
& &  (3.2) \cr}
$$
from which we get (1.13).

We shall now establish the analytic properties of $d^{(p)}$ and $D^{(p)}$.
We begin with the determinant.

Let $\cp_n$ be the collection of partitions 
$\bp = (P_1, \dots P_s)$ of $\{1, \dots, n\}$. We define
$$
T^{(p)}_n = \sum_{\bp \in \cp_n} t_{|P_1|}^{(p)} \dots t_{|P_s|}^{(p)}
\eqno (3.3)
$$
where
$$
t^{(p)}_r = \cases{0 & $\;\;\;\;\;$ if $r<p$ 
\cr
c^{r} \left( {p-2 \over p-1} r \right)! \; \n K \n_r^r & 
$\;\;\;\;\;$ otherwise \cr}
\eqno (3.4)
$$
It follows from (1.16) - (1.18) that
$$
|d^{(p)}_n(K)| \le T^{(p)}_n  \eqno (3.5)
$$

Let $k_j = \# \{P_i: \; |P_i| = j , i =1, \dots, s\}$. Then (3.3) can be 
written as
$$
T^{(p)}_n = n! \sum_{k_1, \dots, k_n} {1 \over k_1 ! \cdots k_n !} 
\left( {t_1^{(p)} \over 1 !} \right)^{k_1} \dots \left( {t^{(p)}_n \over n!}
\right)^{k_n}  \eqno (3.6)
$$
where the summation is over all non negative integers $k_1, \dots, k_n$, 
such that $k_1 + 2 k_2 + \cdots + n k_n = n$.

Notice that
$$
\sum_{n \ge 0} {1 \over n!} \; T_n^{(p)} = \exp \biggl\{ \sum_{j \ge 1}
{1 \over j!} \; t^{(p)}_j  \biggr\}    \eqno (3.7)
$$
can be used with (1.4) and (3.5) to obtain that 
$$
|d^{(p)}(K;\lambda)| \le e^{\eta}    \eqno (3.8)
$$
with 
$$
\eta =  \sum_{j \ge p} {c^{j} \over j!} \left( {p-2 \over p-1} j
\right) ! \; \n \lambda K \n_j^j   \eqno (3.9)
$$
which establishes Theorem 1.1 for $d^{(p)}$.

The proof of (1.22) requires the following Lemma:

\sp
\proclaim Lemma 3.1. Given $p > 2$, the Weierstrass primary factor
satisfies the upper bound
$$
|E(z;p-1)| \le e^{\kappa |z|^{p}} \eqno (3.10)
$$
with 
$$
\kappa = {p-1 \over p} \exp \biggl\{- {p-2 \over 4 p} \biggl[ 
1+ \biggl(1+ 2 \biggl(1+ {\rm cosec} {\pi \over p} \biggr)^{-1} \biggr)^{p-1}
\biggr]^{-1} \biggr\} \;\;.  \eqno (3.11)
$$

\sp

Defferring the proof of Lemma 3.1 to the end of this section, it follows 
from (1.21) and (3.10) that
$$
|d^{(p)}(K;1)| \le \exp \biggl\{\kappa \sum_{i \in I} |\gamma_i|^p \biggr\} 
$$
which establishes (1.22) since by Weyl's inequality$^{[14]}$
$$
\sum_{i \in I} |\gamma_i|^p \le \n K \n_p^p.
$$

To prove convergence of $D^{(p)}$'s series we base on Simon's ideas$^{[6]}$.

Let $\phi \com \psi \in \ch$ be such that $\n \phi\n_2 = \n \psi \n_2 = 1$
and let $B = (\phi \com \cdot) \psi$. We set $A = \mu B + \lambda K$
in Lemma 2.3. Then we have
$$
\eqalignno{
\left[ {d\over d\mu} d^{(p)}(A;1) \right]_{\mu=0} & = 
\left[ \ct_1(R K_{\{1\}}) d^{(p)}(A;1) \right]_{\mu=0} &
\cr
& = \left[ \ct_1(R(K;\lambda) \; B) + \ct_1(\xi_p(\lambda K) \; B) \right] 
d^{(p)}(K;\lambda) & (3.12) \cr}
$$

Since $\ct_1(R(K;\lambda) B) = (\phi, R(K;\lambda) \psi)$, (1.2) and (3.12) 
imply that
$$
(\phi \com D^{(p)}(K;\lambda)\psi) = \left[ {d\over d\mu} d^{(p)}(A;1)
\right]_{\mu=0} - \ct_1(\xi_p(\lambda K) B) d^{(p)}(K;\lambda) \eqno (3.13)
$$

The first term in the right hand side of (3.13) can be estimated as in [6]
$$
\left| {d\over d\mu} d^{(p)}(A;1) \right|_{\mu=0} \le \sup_{|\mu|=1}
|d^{(p)}(A;1)| \le e^{\kappa \n B+\lambda K\n_p^p} \eqno (3.14)
$$
If we use $ \ct_1(|B|^k) \le \n \phi \n_2^k \; \n \psi \n_2^k = 1$
we obtain 
$$
\n B + \lambda K \n_p^p \le \sum_{n=0}^p {p \choose n} |\lambda|^n \; 
\ct_1(|K|^n |B|^{p-n}) \le (1 + |\lambda| \n K \n_p)^p  \eqno (3.15)
$$

The second term of (3.13) can be estimated by using
$$
|(\phi \com \xi_p(\lambda K) \psi)| \le \n \xi_p(\lambda K) \n_\infty
\le \exp \biggl\{ \sum_{j=1}^{p-2} |\lambda|^j \n K \n_p^j \biggr\}
\eqno (3.16)
$$
and the estimate (1.22).

>From (3.13) - (3.16) 
$$
|(\phi \com D^{(p)}(K;\lambda) \psi)| \le 2 \; e^{\kappa 
(1 +\lambda \n K \n_p)^p} \eqno (3.17)
$$
which implies that $D^{(p)}$ is an entire bounded operator valued function
on $\lambda$, proves (1.23) and concludes the proof of theorems 1.1 and 1.4.
\sp
\ni
{\bf Proof of Lemma 3.1}: Given $p \in \{3, 4, \dots\}$ we define a family 
of real valued functions on $\bc$ indexed by $\kappa > 0$ given by
$$
\eqalignno{
\f(\rho,\vf) & \equiv |E(z= (\rho,\vf); p-1)|^2  \; e^{- 2 \kappa \rho^p} & 
\cr
& = (1 + \rho^2 - 2\rho \cos \vf) \exp 
\biggl\{2 \biggl[ \sum_{j=1}^{p-1}
{ \rho^j \over j} \cos j \vf - \kappa \rho^p \biggr] \biggr\} &(3.18) \cr}
$$

Lemma 3.1 is implied if there exists $\kappa = \kappa(p)$ such that
$$
|\f(\rho,\vf)| \le 1  \eqno (3.19)
$$
for any $(\rho, \vf) \in [0,\infty] \times (-\pi, \pi]$. We let ${\cal K}$ be
the set of $\kappa$'s satisfying (3.19). In the sequel we will construct a non
empty set ${\cal M} \ss {\cal K}$ from which (3.11) follows by taking ${\bar
\kappa} = \inf_\kappa {\cal M}$.

One can easily check the following properties of $\f$:

\item{(i)} $\f(0, 0) =1$

\item{(ii)} $\f(\rho, \vf) \ge 0$

\item{(iii)} $\f(\rho, \vf) \longrightarrow 0$ as $\rho \to \infty$

>From these we conclude that (3.19) is violated only if a
non-trivial maximum is developed. 

We first fix $\vf$ and minimize $\f$ with respect to $\rho$. We have 
$$
{d \over d\rho} \f(\rho,\vf) = 2 \biggl[ {(\rho - \cos \vf) \over 
(1 + \rho^2 -2 \rho \cos \vf)} 
+ \biggl( \sum_{j=0}^{p-2} \rho^j \cos (j+1)\vf - p \kappa
\rho^{p-1} \biggr) \biggr] f(\rho,\vf) \eqno (3.20)
$$
which can be written, using $2 \cos (j+1)\vf \; \cos \vf = \cos (j+2)\vf +
\cos j \vf$, as
$$
{ -2p\kappa \rho^{p-1} \over 1 + \rho^2 - 2\rho \cos \vf } \left[ \rho^2 - 
\left(
2 \cos \vf + {1 \over p \kappa} \cos (p-1) \vf \right) \rho + 1 +
{1 \over p \kappa} \cos p \vf \right] f(\rho,\vf) \eqno (3.21)
$$

Thus, non-trivial solutions of ${d\over d\rho} \f(\rho,\vf) = 0$,
$$
\rho_\pm (\vf) = \cos \vf + {1 \over 2p\kappa} \cos (p-1)\vf \pm \Delta
\com \eqno (3.22)
$$
exist provided
$$
\Delta^2 = \left( \cos \vf - {1\over 2p\kappa} \cos (p-1)\vf \right)^2 + 
{1\over p\kappa} \cos (p-2)\vf -1 \ge 0 \eqno (3.23)
$$


We notice that, since
$$
{d^2 \over d \rho^2} \f(\rho_\pm ,\vf) = \mp {4 p \kappa \rho^{p-1}_\pm 
\Delta^2
\over 1 + \rho_\pm^2 - 2 \rho_\pm \cos \vf} f(\rho_\pm, \vf) \eqno (3.24)
$$
$\rho_+$ ($\rho_-$) is a maximum (minimum) of $\f$ for each direction $\vf$.
Moreover, if ${\pi \over p} < |\vf| < (p-1) {\pi \over p}$ and $\kappa >
{1 \over 2p} (1+ {\rm cosec} {\pi \over p} ) \equiv \kappa_1$, we have 
$\Delta^2 < 0$ and if $|\vf| \ge (p-1) {\pi \over p}$ and $\kappa >\kappa_1$, 
we have $\rho_{\pm} < 0$.

We now fix $\rho$ and use the trigonometric relation $2 \sin j \vf \; 
\cos  \vf = \sin (j+1) \vf + \sin (j-1) \vf$ to get
$$
{d \over d\vf} \f(\rho, \vf) = {2 \; \rho^p \over 1 + \rho^2 -2 \rho \cos 
\vf} \left[ \sin p \vf - \rho \sin (p-1) \vf \right] f(\rho, \vf) \eqno
(3.25)
$$

We find that any solution ${\bar \vf} = {\bar \vf}(\rho)$ of 
${d \over d\vf} \f(\rho, \vf) = 0$ satisfies
$$
\sin p {\bar \vf} = \rho \sin (p-1) {\bar \vf}  \eqno (3.26)
$$

Notice that ${\bar \vf} = 0$ satisfies (3.26) and ${d^2 \over d \vf^2} f(\rho, 
{\bar \vf})$ is negative if and only if 
$$
\rho > {p \cos p {\bar \vf} \over (p-1) \cos (p-1) {\bar \vf}}  \eqno (3.27)
$$
which implies that $(\rho_+(0), 0) = (1 + {1\over p \kappa} \com 0)$ is a local
maximum of $f_\kappa$ if $\kappa < 1 - {1\over p} \equiv \kappa_2$.

In fact, from the above analysis we conclude that 
$(1 + {1 \over p \kappa} \com  0)$ is the unique non - trivial maximum 
of $\f$ provided
$$
\rho_+(\vf) > {p \sin p \vf \over (p-1) \sin (p-1) \vf}  \eqno(3.28)
$$
holds for $|\vf| \le {\pi \over p}$ with $\kappa_1 \le \kappa \le \kappa_2$.

Assuming (3.28) valid, we can replace (3.19) by the condition
$$
\f(\rho_+(0),0) \le 1  \eqno (3.29)
$$
which is implied by
$$
\sum_{j=1}^{p-1} {1 \over j} \left(1 + {1 \over p \kappa} \right)^j
\le \kappa \left( 1+ {1 \over p \kappa} \right)^p  
+  \ln p\kappa  \eqno (3.30)
$$

Since 
$$
\eqalignno{
\sum_{j=1}^{p-1} {1\over j} \left(1 + {1\over p\kappa} \right)^j & \le
\left(1+ {1\over p\kappa}\right)  +
\int_{{1\over p}}^{{p-1\over p}} {1\over x} \left(1+ {1\over p\kappa} 
\right)^{px} dx &
\cr
& \le \ln (p-1) + \left( 1 + {1\over p\kappa} \right)^{p-1}
- {1\over 4} \left(1 - {2\over p} \right) & (3.31) \cr}
$$
(3.30) is implied by 
$$
\left(1 - \kappa - {1\over p} \right) \left( 1 +{1 \over p\kappa} 
\right)^{p-1} \le \ln {p \over p-1} \kappa + {1\over 4} \left(1 - {2\over p}
\right) \; .  \eqno (3.32)
$$
It has been used in (3.31) that $\int {e^{a x} \over x} dx = C + \ln a x
+ \sum_{k \ge 1} {(a x)^k \over k\cdot k!}$ and $\kappa < 1 - {1\over p}$.

Now, for any  $\kappa \ge {1 \over 2p} (1+ {\rm cosec} {\pi \over p})$ we have
$$
\left(1-\kappa -{1\over p}\right) \left(1+ {1 \over p\kappa}\right)^{p-1} 
\le \left(-\ln {p\over p-1} \kappa \right) \left[1+ \left( 1+ 
2{\rm cosec} {\pi \over p}\right)^{-1} \right]^{p-1} \eqno (3.33)
$$
and (3.32) is implied by 
$$
\kappa \ge  {p-1 \over p} \exp \biggl\{- {p-2 \over 4 p} \biggl[
1+ \biggl(1+ 2\biggl( 1+ {\rm cosec} {\pi \over p}\biggr)^{-1} \biggr)^{p-1}
\biggr]^{-1} \biggr\} \equiv \kappa_3  \eqno (3.34)
$$
Notice that $\kappa_1 < \kappa_3 < \kappa_2$, which implies that 
${\cal M} = \{\kappa \; : \; \kappa_3 \le \kappa < 
\kappa_2 \} \ss {\cal K}$ is a non-empty set and (3.11) follows.

We conclude the proof of Lemma 3.1 by showing (3.28). (3.23) and (3.34)
imply that $\Delta^2 \le (\cos \vf - 1/2p\kappa \cos(p-1) \vf)^2$ which
can be used with (3.22) and $2 \sin(p-1)\vf \cos (p-2)\vf = \sin \vf
+\sin(p-2)\vf$ to replace (3.28) by 
$$
(p-1) \sin (p-2) \vf \ge \sin \vf  \eqno (3.35)
$$
This concludes our proof since (3.35) is true for $|\vf| \le {\pi \over p}$
provided $p>2$.

\vskip 1in

\ni
{\bf 4. FREDHOLM FORMULA}
\sp

We are here concerned with integral equations of the form (1.1). Our Hilbert
space is $\ch = L_2(\L)$ and $K$ has an integral kernel on $\L \times \L$,
with $\L \subseteq \br^d$ so that, if $K \in \cc_p$ with $p$ even, then
$$
\int \prod_i d^dx_i \; |K(x_1, x_2)K^{\ast}(x_2, x_3) \dots K(x_{p-1}, x_p)
K^{\ast}(x_p, x_1)| < \infty  \eqno (4.1)
$$

We notice that
$$
\ct_n(\wedge^n K) = n! \int \prod_i d^dx_i \sum_\pi (-1)^{|\pi|} 
K(x_1,x_{\pi(1)}) \dots K(x_n,x_{\pi(n)}) \eqno (4.2)
$$
which implies that Fredholm's formula for determinant $d^{(1)}(K)$ is 
just Grothendieck's formula.

A simple derivation of the Fredholm's formula for $D^{(1)}(K)$ can be obtained
from (3.13). By using
$$
d^{(1)}(\mu B+ \lambda K;1) = d^{(1)}(B;\mu) d^{(1)}(R(B;\mu)K;\lambda) 
\eqno (4.3)
$$
and (2.7), (3.13) (with $p=1$) can be written as
$$
\eqalignno{
( \phi, D^{(1)}(K;\lambda ) \psi ) & = \sum_{n \ge 0} {\lambda^n \over n!}
\left[ {d \over d\mu} \left( \ct_n(\wedge^n R(B;\mu ) \cdot \wedge^n K)
d^{(1)}(B;\mu) \right) \right]_{\mu=0} &
\cr
& = \sum_{n\ge 0} {\lambda^n \over n!} \left( \ct_1(B) \ct_n(\wedge^nK) -
\ct_n(d\wedge^n B \cdot \wedge^n K) \right) &
\cr
& = \sum_{n \ge 0} {\lambda^n \over n!} \ct_{n+1}(\wedge^n K \wedge B)
& (4.4) \cr}
$$                                                
which can be immediately recognized as Fredholm's formula after we rewrite
its coefficients $(\phi, D_n^{(1)} \psi) = \ct_{n+1}(\wedge^n K \wedge
B)$ as
$$
(n+1)! \int \prod_i dx_i \sum_\pi (-1)^{|\pi|}
K(x_1, x_{\pi(1)}) \dots K(x_n, x_{\pi(n)}) \phi(x_{n+1}) \psi(x_{\pi(n+1)})
\eqno (4.5)
$$

In our last application we generalize Fredholm's expression of $D^{(p)}$
for $p>1$. It is based on an explicitly calculation of ${d\over d\lambda}
d^{(p)}(A;1)$ with $A = \mu B + \lambda K$ as in (3.12).

Using Lemma 2.3 we have
$$
d^{(p)}(A;1) = \sum_{n \ge 0} {\lambda^n \over n!} {\tilde d}^{(p)}_n(\mu) 
\eqno (4.6)
$$
where
$$
{\tilde d}^{(p)}_n(\mu) = \sum_{\bp \in \cp_n} \ct_s(\wedge^s R(B;\mu)
\cdot \tks) \; d^{(p)}(B;\mu)  \eqno (4.7)
$$
and
$$
\eqalignno{
{\widetilde K}_Q & = \left[ \left( {d \over d\lambda} + \xi_p(\mu B + 
\lambda) \right)^{|Q|} (1 + \mu B + \lambda) \right]_{\lambda = 0} 
K^{|Q|} & \cr
& \equiv {\widetilde C}^{(p)}_{|Q|}(\mu) \; K^{|Q|} & (4.8) \cr}
$$

Notice that ${\widetilde C}^{(p)}_k(0) = C^{(p)}_k$ as defined by (1.14).

Since ${d \over d\mu} d^{(p)}(B;\mu) |_{\mu=0} = 0$ for any $p >1$, 
we have
$$
\left[{d \over d\mu} {\tilde d}^{(p)}_n(\mu) \right]_{\mu =0} =
\sum_{\bp \in \cp_n} \left( {\cwt}_\bp - {\cyt}_\bp \right) \eqno (4.9)
$$
where
$$
\eqalignno{
{\cwt}_\bp & = {d \over d\mu} \left[ \ct_s(\tks) \right]_{\mu=0} &
\cr
& = \sum_j \left[{1\over B} {d \over d\mu} {\widetilde C}^{(p)}_{|P_j|} 
\right]_{\mu=0}
\left( \prod_{i \not = j} C^{(p)}_{|P_i|} \right) \ct_s(K^{|P_1|} 
\wedge \cdots \wedge B K^{|P_j|} \wedge \cdots \wedge K^{|P_s|}) &
(4.10) \cr}
$$
and
$$
{\cyt}_\bp = \left( \prod_i C^{(p)}_{|P_i|} \right) \ct_s(d\wedge^s B \cdot 
K^{|P_1|} \wedge \cdots \wedge K^{|P_s|}) \eqno (4.11)
$$


We now notice that
$$
\left[ {1\over B} {d\over d\mu} {\widetilde C}^{(p)}_k \right]_{\mu=0}
- C^{(p)}_k = C^{(p)}_{k+1}  \eqno (4.12)
$$

>From (4.6) - (4.12) and Lemma 2.1 we conclude that  
$$
{d\over d\mu} d^{(p)}(A;1) =
\sum_{n \ge 0} {\lambda^n \over n!} \Delta^{(p)}_n  \eqno (4.13)
$$
where
$$
\Delta_n^{(p)} = \sum_{\bp \in \cp_n} C^{(p)}_{|P_1|+1} \left(
\prod_{j=2}^s C^{(p)}_{|P_j|} \right) \ct_s(d\wedge^s B \cdot K^{|P_1|}
\wedge \cdots \wedge K^{|P_s|})
\eqno (4.14)
$$
This expression can be expanded as in (4.5) to exhibit its determinant form.

Our final expression is obtained by combining (3.13), (4.13) and (4.14).

\vskip 1in
\ni
{\bf ACKNOWLEDGEMENTS}

I would like to thank T. Hurd and G. Slade for the hospitality at 
McMaster University. I also thank T. Hurd and P. Faria da Veiga for
many helpful discussions.

\vskip 1in
\ni
{\bf REFERENCES}

\sp
\item{[1]} I. Fredholm, ``Sur une classe d'\'{e}quation fonctionelle",
Acta Math. {\bf 27}, 365 - 390 (1903).

\item{[2]} H. Poincar\`{e}, ``Remarques diverses sur l'\'{e}quation de
Fredholm", Acta Math. {\bf 33}, 57 - 86 (1910).

\item{[3]} J. Plemelj, ``Zur theorie Fredholmshen funktionalgleichung",
Monat. Math. Phys. {\bf 15} 93 - 128 (1907).

\item{[4]} F. Smithies, ``The Fredholm theory of integral equations",
Duke Math. J. {\bf 8}, 107 - 130 (1941).

\item{[5]} H. J. Brascamp, ``The Fredholm theory of integral equations
for special types of compact operators on separable Hilbert space",
Comp. Math. {\bf 21}, 59 - 80 (1969).

\item{[6]} B. Simon, ``Notes on infinite determinants of Hilbert space
operators", Adv. Math. {\bf 24}, 244 - 273, (1977).

\item{[7]} B. Simon, ``Trace ideals and their applications", London
Math. Soc. Lecture Note {\bf 35} (1979).

\item{[8]} see e.g. R. Courrant and D. Hilbert, ``Methods of Mathematical
Physics", John Wiley (1989).

\item{[9]} A. Grothendieck, ``La th\'{e}orie de Fredholm", Bull. Soc. Math.
(France) {\bf 84}, 319 - 384 (1956).

\item{[10]} A. Cooper and L. Rosen, ``The weakly coupled Yukawa$_2$ field
theory: cluster expansion and Wightman axioms", Trans. Amer. Math. Soc.
{\bf 234} 1 - 88 (1977).

\item{[11]} T.R. Hurd, D.H.U. Marchetti and P.A.F. da Veiga, "On the
mass generation in the Gross- Neveu$_2$  quantum
field theory", work in preparation.

\item{[12]} R. Nevanlinna, "Analytic Function", Springer - Verlag 
(1970).

\item{[13]} V. B. Lidskii, "Non - selfadjoint operators with a trace",
Dokl. Akad. SSSR {\bf 125}, 485 - 587 (1959).

\item{[14]} H. Weyl, " Inequalities between two kinds of eigenvalues
of a linear transformation", Proc. Nat. Acad. Sci. U.S.A. {\bf 35}, 
408 - 411 (1949).

\end